\documentclass[%
reprint,
amsmath,
amssymb,
aps,
twocolumn,
]{aastex61}

\usepackage{graphicx}
\usepackage{dcolumn}
\usepackage{bm}
\usepackage{color}
\usepackage{natbib}
\begin{document}

\newcommand{\oh}{H$\beta$+[\hbox{{\rm O}\kern 0.1em{\sc iii}}]}
\newcommand{\lam}{[\hbox{{\rm O}\kern 0.1em{\sc iii}}]$\lambda 5007$}
\newcommand{\OIII}{[\hbox{{\rm O}\kern 0.1em{\sc iii}}]}
\newcommand{\prospector}{\hbox{{\rm P}\kern0.1em{\sc rospector}}}
\newcommand{\zfourge}{\hbox{\kern0.1em{\sc zfourge}}} 
\newcommand{\numSFG}{167}
\newcommand{\numEELG}{19}
\newcommand{\emcee}{\texttt{emcee}}
\newcommand{\dfour}{\textit{D}(4000)}
\newcommand{\forrest}{Forrest et al.~(2018, accepted; hereafter F18)}
\newcommand{\fshort}{F18}
\newcommand{\lshort}{L17}
\newcommand{\tran}{Tran et al.~2018, in prep.}
\newcommand{\ptran}{Tran et al.~(2018, in prep.)}
\newcommand{\lejaprep}{Leja et al.~(in prep.)}

\shortauthors{Cohn et al.}

\title{~\zfourge: Extreme 5007 \AA\ emission may be a common early-lifetime phase for star-forming galaxies at $\MakeLowercase{z}>2.5$}

\author[0000-0003-1420-6037]{Jonathan H. Cohn}
\affil{George P. and Cynthia W. Mitchell Institute for Fundamental Physics and Astronomy, Department of Physics \& Astronomy, Texas A\&M University, College Station, TX 77843, USA}

\author[0000-0001-6755-1315]{Joel Leja}
\affil{Harvard-Smithsonian Center for Astrophysics, 60 Garden Street, Cambridge, MA 02138, USA}
\affil{NSF Astronomy and Astrophysics Postdoctoral Fellow}

\author[0000-0001-9208-2143]{Kim-Vy H. Tran}
\affiliation{School of Physics, University of New South Wales, Kensington, Australia}
\affiliation{George P. and Cynthia W. Mitchell Institute for Fundamental Physics and Astronomy, Department of Physics \& Astronomy, Texas A\&M University, College Station, TX 77843, USA}
\affiliation{Australian Astronomical Observatory}
\affiliation{ARC Centre for Excellence in All-Sky Astrophysics in 3D (ASTRO 3D)}

\author[0000-0001-6003-0541]{Ben Forrest}
\affiliation{George P. and Cynthia W. Mitchell Institute for Fundamental Physics and Astronomy, Department of Physics \& Astronomy, Texas A\&M University, College Station, TX 77843, USA}
\affiliation{Department of Physics and Astronomy, University of California, Riverside, CA 92521, USA}

\author[0000-0002-9280-7594]{Benjamin D. Johnson}
\affiliation{Harvard-Smithsonian Center for Astrophysics, 60 Garden Street, Cambridge, MA 02138, USA}

\author{Megan Tillman}
\affiliation{George P. and Cynthia W. Mitchell Institute for Fundamental Physics and Astronomy, Department of Physics \& Astronomy, Texas A\&M University, College Station, TX 77843, USA}

\author[0000-0002-2250-8687]{Leo Alcorn}
\affiliation{George P. and Cynthia W. Mitchell Institute for Fundamental Physics and Astronomy, Department of Physics \& Astronomy, Texas A\&M University, College Station, TX 77843, USA}
\affiliation{LSSTC Data Science Fellow}

\author[0000-0002-1590-8551]{Charlie Conroy}
\affiliation{Harvard-Smithsonian Center for Astrophysics, 60 Garden Street, Cambridge, MA 02138, USA}

\author[0000-0002-3254-9044]{Karl Glazebrook}
\affiliation{Centre for Astrophysics and Supercomputing, Swinburne University of Technology, Hawthorn, Victoria 3122, Australia}

\author{Glenn G. Kacprzak}
\affiliation{Centre for Astrophysics and Supercomputing, Swinburne University of Technology, Hawthorn, Victoria 3122, Australia}

\author{Daniel D. Kelson}
\affiliation{The Observatories of the Carnegie Institution for Science, 813 Santa Barbara Street, Pasadena, CA 91101, USA}

\author[0000-0003-2804-0648]{Themiya Nanayakkara}
\affiliation{Leiden Observatory, Leiden University, PO Box 9513, 2300 RA Leiden, The Netherlands}

\author[0000-0001-7503-8482]{Casey Papovich}
\affiliation{George P. and Cynthia W. Mitchell Institute for Fundamental Physics and Astronomy, Department of Physics \& Astronomy, Texas A\&M University, College Station, TX 77843, USA}

\author[0000-0002-8282-9888]{Pieter G. van Dokkum}
\affiliation{Astronomy Department, Yale University, New Haven, CT 06511, USA}

\author[0000-0002-9211-3277]{Tiantian Yuan}
\affiliation{Centre for Astrophysics and Supercomputing, Swinburne University of Technology, Hawthorn, Victoria 3122, Australia}
\affiliation{ARC Centre for Excellence in All-Sky Astrophysics in 3D (ASTRO 3D)}
\affiliation{Research School of Astronomy and Astrophysics, The Australian National University, Cotter Road, ACT 2611, Australia}

\correspondingauthor{Jonathan Cohn}
\email{joncohn@tamu.edu}

\submitjournal{ApJ}

\begin{abstract}
Using the \prospector\ spectral energy distribution (SED) fitting code, we analyze the properties of \numEELG\ Extreme Emission Line Galaxies (EELGs) identified in the bluest composite SED in the \zfourge\ survey at $2.5 \leq z \leq 4$. \prospector\ includes a physical model for nebular emission and returns probability distributions for stellar mass, stellar metallicity, dust attenuation, and nonparametric star formation history (SFH). The EELGs show evidence for a starburst in the most recent 50 Myr, with the median EELG having a specific star formation rate (sSFR) of 4.6 Gyr$^{-1}$ and forming 15\% of its mass in this short time. For a sample of more typical star-forming galaxies (SFGs) at the same redshifts, the median SFG has a sSFR of 1.1 Gyr$^{-1}$ and forms only $4\%$ of its mass in the last 50 Myr. We find that virtually all of our EELGs have rising SFHs, while most of our SFGs do not. From our analysis, we hypothesize that many, if not most, star-forming galaxies at $z \geq 2.5$ undergo an extreme \oh\ emission line phase early in their lifetimes. In a companion paper, we obtain spectroscopic confirmation of the EELGs as part of our {\sc MOSEL} survey. In the future, explorations of uncertainties in modeling the UV slope for galaxies at $z>2$ are needed to better constrain their properties, e.g. stellar metallicities.

\end{abstract}

\section{\label{intro}Introduction}
Galaxies with strong ionizing emission at $z \geq 6$ are thought to be responsible for driving reionization (e.g. \citealt{Nakajima2014}). Such systems are expected to have extreme emission lines with equivalent widths (EWs) of hundreds to over $1000$ \AA\ \citep{Wel2011,Maseda2013,Maseda2014,Amorin2015,Forrest2017}, stellar masses $\lesssim 10^{9}$ M$_{\odot}$ \citep{Maseda2014,Forrest2017}, sub-solar metallicities \citep{Wel2011,Amorin2015}, and specific star formation rates (sSFRs) up to $10 - 100$ Gyr$^{-1}$ \citep{Wel2011,Maseda2014,Amorin2015,Forrest2017}.

Extreme emission line galaxies (EELGs) are rare in the local universe (e.g. green peas, \citealt{Cardamone2009}; luminous compact galaxies, \citealt{Izotov2011}), but the number density of EELGs increases with increasing redshift (e.g. \citealt{Ouchi2009,Labbe2013,Smit2014,Holden2016}). Based on colors and EWs, \citet{Smit2015} found that as many as $50$\% of star-forming galaxies (SFGs) are EELGs at $z\sim 7$. As such, EELGs may be the progenitors of more typical SFGs at $z \sim 2$, when the global peak in star formation density in the Universe took place (e.g. \citealt{Madau1998,Sobral2013}).

Several studies show that EELGs have bursting (e.g. \citealt{Atek2011,Wel2011,Maseda2013}) or rising (e.g. \citealt{Sanders2016}) star formation rates (SFRs) with high sSFRs and young stellar ages \citep{Forrest2017}. However, studies have yet to fully explore the star formation {\it histories}. A better understanding of the SFHs provides more accurate estimates of galaxy properties (e.g. stellar mass, age, and metallicity) and insight into the duration of the extreme emission, i.e. the duty cycle. Additionally, modeling nebular emission accurately --- especially when the emission comprises a significant fraction of the broadband photometric flux --- remains a challenge.

Analyzing the SFHs and other properties of high-redshift EELGs requires photometry capable of detecting the faint continuum wavelengths outside of emission lines. The \zfourge\ survey meets these requirements with deep multi-wavelength photometry, including as many as 40 measurements in the rest frame UV to near-IR for galaxies at $1 \leq z \leq 4$ \citep{Straatman2016}. Accurately modeling these EELGs also calls for state-of-the-art stellar population synthesis models, as well as Bayesian statistics to disentangle the effects of SFH and other galaxy parameters, such as metallicity. This type of analysis is now possible with the \prospector\ SED fitting code (\citealt{Leja2017}; hereafter \lshort).

In this paper, we investigate the properties of \numEELG\ \oh\ candidate EELGs at $2.5 \leq z \leq 4$ selected in \forrest\ in composite spectral energy distributions (SEDs) in the \zfourge\ survey. We create a model based on \prospector-$\alpha$ (\lshort) to estimate the stellar masses, metallicities, dust attenuation, and SFHs for our EELG sample and for a sample of \numSFG\ more typical SFGs in the same redshift range. Ours is the first study to use \prospector\ at $z \gtrsim 2$, and we are the first to compare SFHs for different galaxy populations using composite SEDs. We detail our galaxy populations and describe our use of \prospector\ in \S \ref{methods}, display our results in \S \ref{results}, and summarize our conclusions in \S \ref{conclusions}.

\section{\label{methods}Methods}
\subsection{\label{data}Data}
Our data are from the FourStar Galaxy Evolution (\zfourge) Survey, a deep, near-IR, medium-band survey (see \citealt{Straatman2016} for more information) completed with the FourStar instrument on the Magellan telescope in the \textit{Chandra} Deep Field South (CDFS; \citealt{Giacconi2002}), Cosmic Evolution Survey field (COSMOS; \citealt{Scoville2007}), and Ultra Deep Survey field (UDS; \citealt{Lawrence2007}). With extensive rest frame wavelength coverage for galaxies at $1 \leq z \leq 4$, the survey yields exceptionally accurate photometric redshifts ($1-2\%$; \citealt{Nanayakkara2016}).

Taking advantage of these accurate photometric redshifts and well-sampled SEDs, \fshort\ uses EAZY to generate 22 rest-frame fluxes for each galaxy and subsequently groups them by similarities in the resultant colors.
Groups of galaxies with similar SED shapes are then de-redshifted and scaled to create low resolution spectra, or composite SEDs, revealing galaxies with a wide array of SED shapes.
Notably, strong emission features such as H$\alpha$ and \oh\ are measurable.
In \fshort\ this is done for $\sim 7000$ galaxies from $1<z<4$ with \textit{SNR$_K$}$>20$.

In this work we focus on galaxies from two of these composite SEDs over $2.5 \leq z \leq 4$. The composite SED with the largest \oh\ EW ($\sim 2600$ \AA) and steepest UV slope ($-2.05$; i.e. the bluest composite SED) measured in \fshort, containing \numEELG\ galaxies, is selected as our EELG sample. Four of these EELGs have spectroscopically-confirmed redshifts from the {\sc MOSEL} survey (\tran). In {\sc MOSEL}, we find that the \fshort\ photometrically-selected emission-line galaxies trace out the same property distributions as the spectroscopically-confirmed sample.

\begin{figure}
\includegraphics[width=0.47\textwidth]{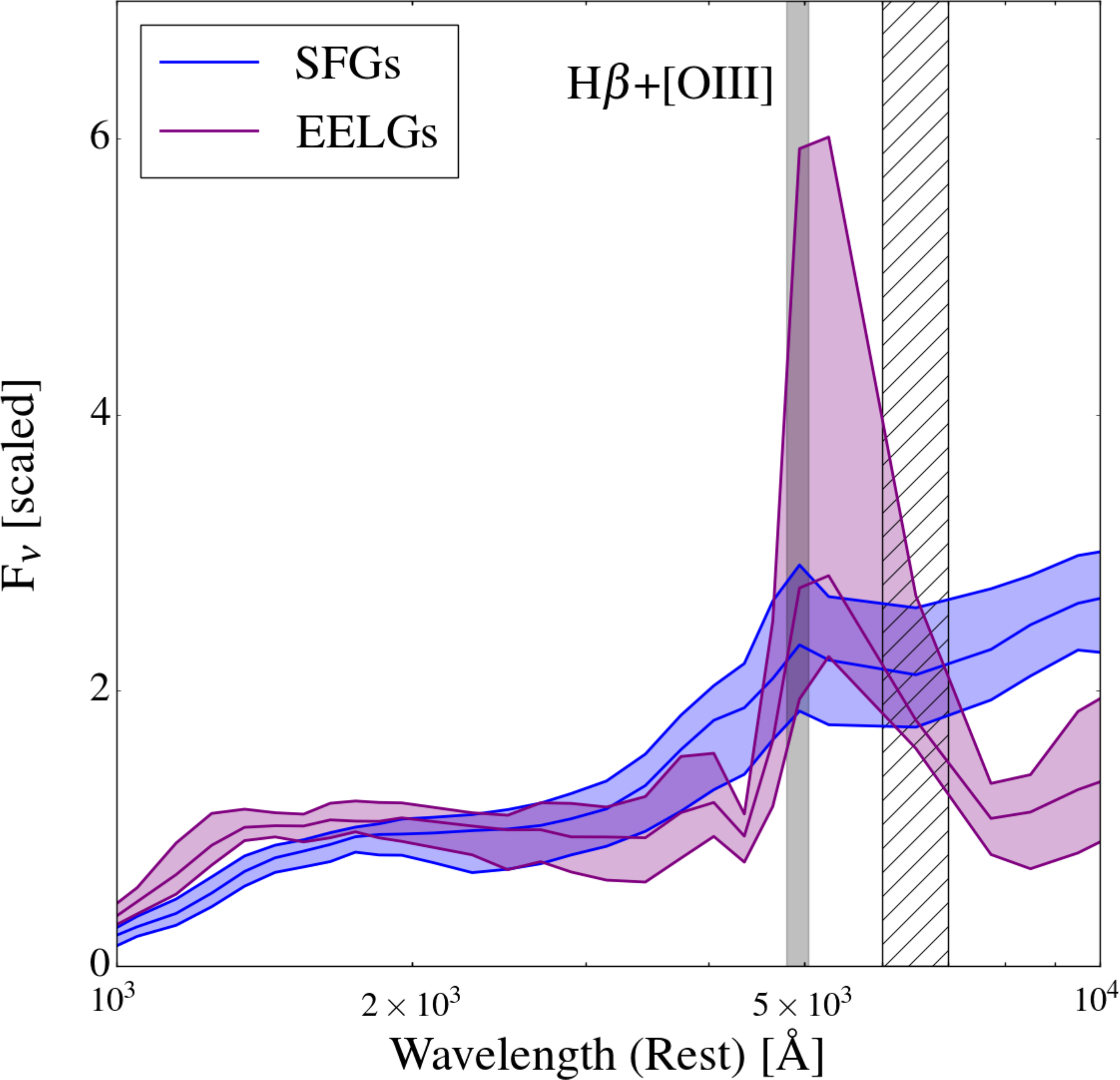}
\caption{We plot the median and 68 percentiles of the observed photometric rest-frame SEDs for galaxies in our EELG (purple) and SFG (blue) samples, highlighting the observed differences --- namely the blue UV slopes and strong optical emission lines for our EELGs --- that are used to select the samples in \fshort. The vertical gray-shaded area shows the \oh\ emission line region. The hashed region shows the approximate wavelengths where we generally lack rest-frame photometry for our galaxies. We note that $\sim 50$\% of our EELGs are not detected red-ward of $\sim 1\mu$m.}
\label{seds}
\end{figure}

We choose the composite SED with the largest number of analogs (\numSFG\ galaxies) to represent typical SFGs at $2.5 \leq z \leq 4$. This composite SED has less extreme emission (\oh\ EW $\sim 200$ \AA) and a more moderate UV slope ($-1.52$), as determined in \fshort. In Figure \ref{seds}, we plot distributions of the observed photometric SEDs for the galaxies in each of our samples, displaying the differences in SED shapes between the samples.

\subsection{\label{model}Model}
\begin{figure*}
\includegraphics[width=\textwidth]{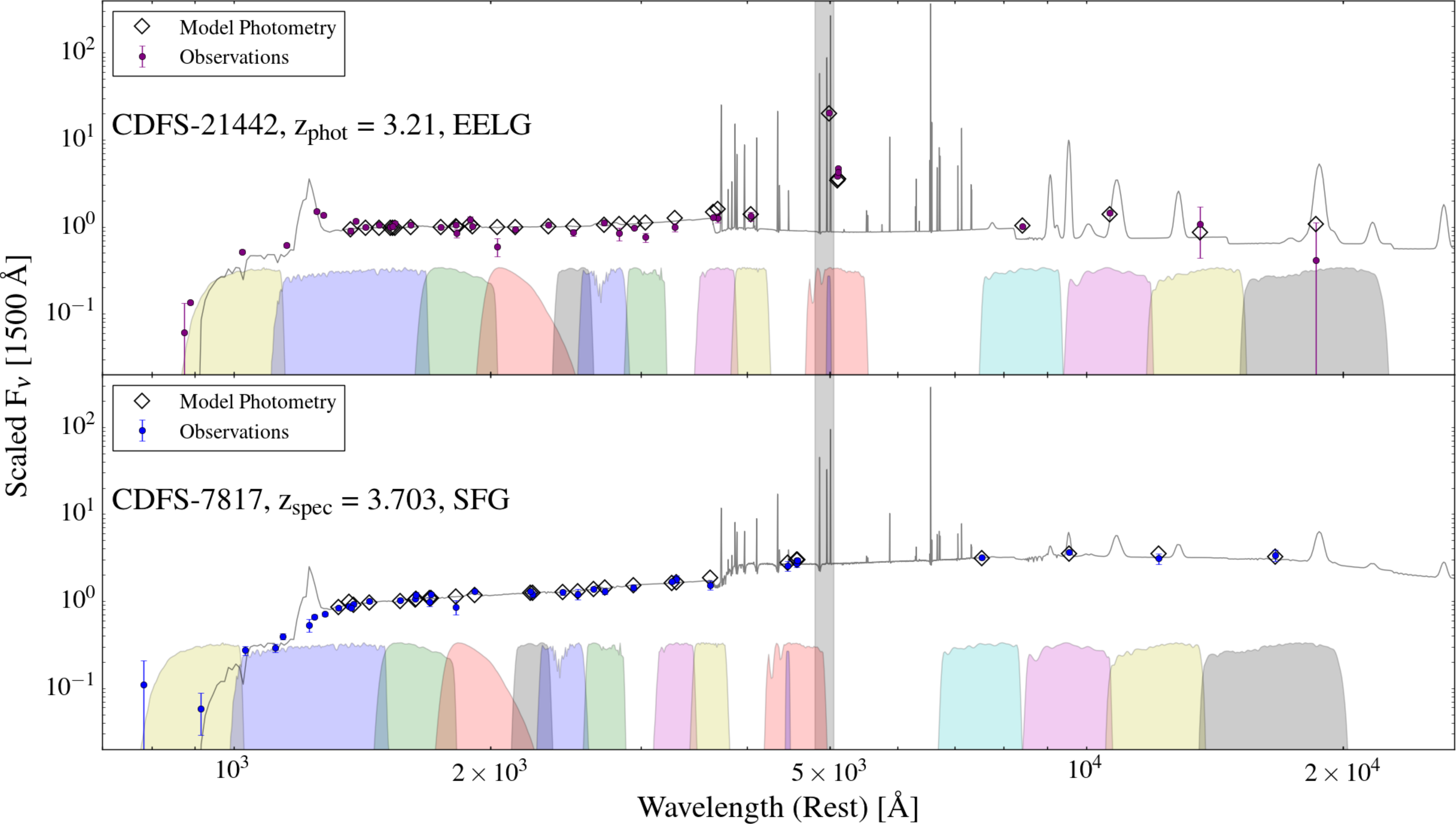}
\caption{Top: rest frame SED for the EELG CDFS-21442, with observations plotted in purple. Bottom: rest frame SED for the SFG CDFS-7817, with observations in blue. The model photometry (black diamonds) and model spectrum output by \prospector\ are also plotted. The observations in filters contaminated by Lyman-$\alpha$ emission are masked in the model fits, as are observations blue-ward of Lyman-$\alpha$. Overlaid is a representative subset of filter curves (see \citet{Straatman2016} for the full set of filter curves). \prospector\ fits stronger emission (see \oh\ region, shaded) and a more negative UV slope to the EELG.}
\label{filters}
\end{figure*}

\begin{figure*}
\includegraphics[width=\textwidth]{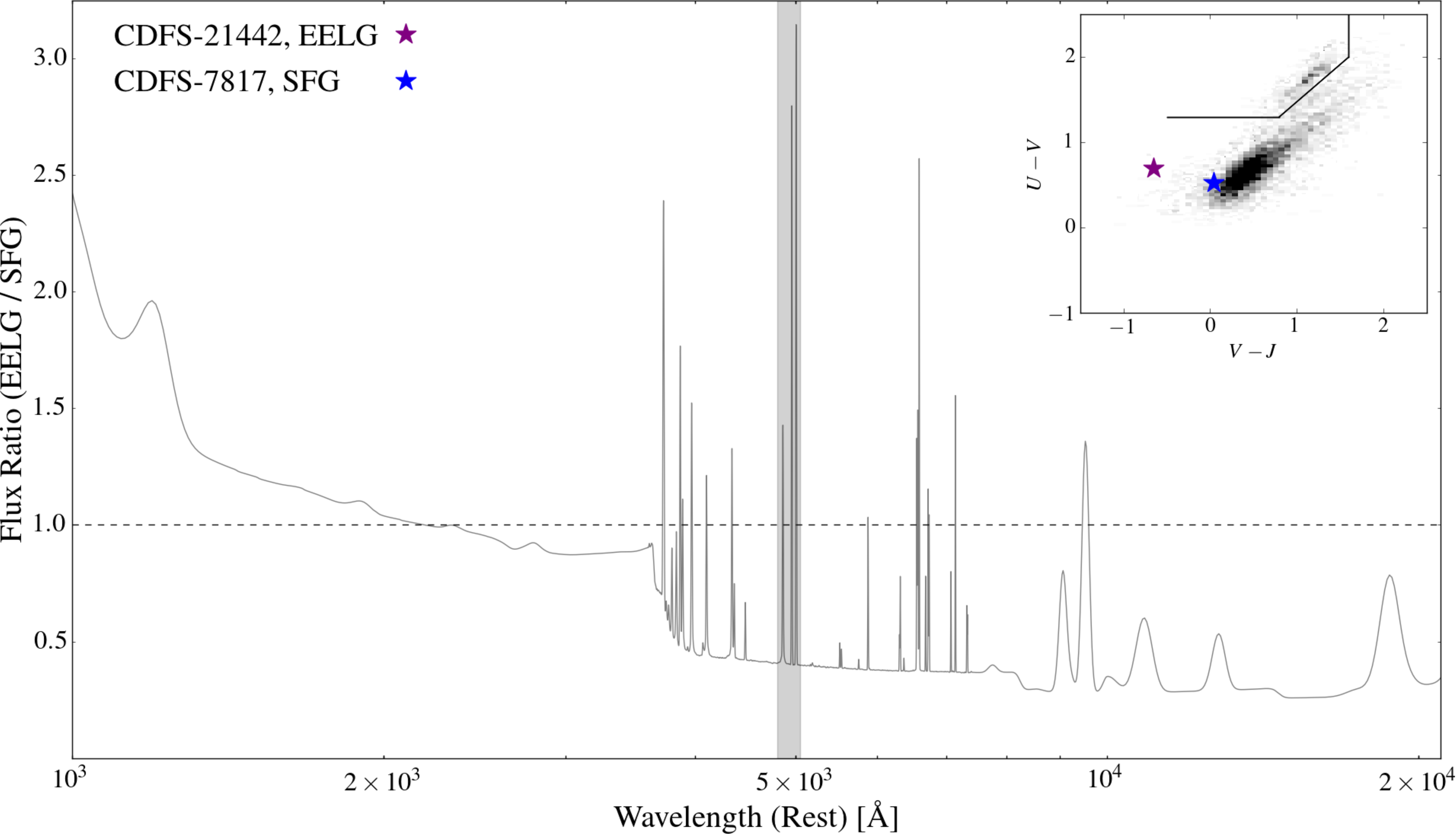}
\caption{We take the ratio of the spectra fit by \prospector\ to the EELG CDFS-21442 and the SFG CDFS-7817, highlighting the differences in their shapes. The EELG shows more UV flux, an excess of emission line flux (especially \oh; shaded region), and a dearth of NIR flux compared to the SFG. These differences are typical for our EELGs and SFGs. The inset UVJ plot displays the position of all CDFS, COSMOS, and UDS galaxies in \zfourge\ in grayscale. The black line on the UVJ diagram separates quiescent (above the line) and star-forming galaxies (below), while the positions of CDFS-21442 and CDFS-7817 are highlighted with a purple and blue star, respectively.}
\label{ratio}
\end{figure*}

We use \prospector\footnote{doi:10.5281/zenodo.1116491}, a python-based SED fitting code, with a model based on \prospector-$\alpha$ (see \lshort), to derive galaxy properties from photometry. \prospector\ uses the Flexible Stellar Population Synthesis package (FSPS; \citealt{Conroy2010}, \citealt{Conroy2009}) via python-FSPS\footnote{doi:10.5281/zenodo.12157}, taking into account dust attenuation, re-radiation, and nebular emission \citep{Byler2017}. The code uses a Bayesian inference framework and samples the parameter space with \emcee\ \citep{Mackey2013}. We employ MESA Isochrones \& Stellar Tracks (MIST; see \citealt{Dotter2016}, \citealt{Choi2016}, \citealt{Paxton2011, Paxton2013, Paxton2015}) and a \citet{Calzetti1994} dust law and adopt a \citet{Chabrier2003} IMF and a WMAP9 cosmology \citep{Hinshaw2013}.

\subsubsection{SED parameters}
Our fits include 9 free parameters: the stellar mass, stellar metallicity, gas-phase metallicity, dust attenuation, and 5 independent nonparametric SFH bins. The stellar and gas-phase metallicities are fit separately in order to allow the model to reproduce the extreme nebular conditions required to drive extraordinary \oh\ emission. We describe the SFH parameters in more depth in \S \ref{npsfh}. For more information on the free parameters we use in \prospector, we refer the reader to \lshort.

For our fits, we turn on nebular and dust emission and fix the ionization parameter to $\log_{10}(U) = -1.0$. Bands contaminated by Lyman-$\alpha$ emission are masked, as the line is difficult to model accurately due to the effects of resonant scattering (e.g. \citealt{Erb2014}). \prospector\ accounts for IGM absorption following \citet{Madau1995}, but we note that this is not critical because we mask photometry blue-ward of Lyman-$\alpha$. \prospector\ finds the full PDF for each parameter and generates the spectrum and photometry associated with each parameter vector.

Figure \ref{filters} displays the \prospector-derived best-fit model spectrum and SED along with observed photometry for an EELG and SFG in CDFS. The figure also shows a subset of filters in CDFS, displaying our extensive rest frame wavelength coverage. The difference in SED shape that \prospector\ fits to these galaxies is displayed in Figure \ref{ratio}, highlighting the steep UV slope of the EELG relative to the SFG. The EELG also exhibits excess emission line flux compared to the SFG, particularly in \lam. These differences are largely consistent across the EELGs and SFGs in our sample (see Figure \ref{seds}).

\subsubsection{Comparison to FAST\label{fast}}
We have used FAST \citep{Kriek2009} in past studies (e.g. \citealt{Forrest2017}) to estimate parameters of EELGs, but note that FAST was originally tailored for studying massive galaxies at $z \sim 2$. There are several differences between FAST and \prospector: FAST holds stellar metallicity fixed, uses $\tau$-based SFH models that do not allow for bursts, and uses stellar population synthesis models from \citet{Bruzual2003}\footnote{BC03 uses Padova and Geneva stellar evolutionary tracks} \citep{Kriek2009}. The \citet{Forrest2017} FAST fits also include emission line templates from \citet{Salmon2015}.

In contrast, \prospector\ allows stellar and gas-phase metallicity to vary and fits for nonparametric and thus more diverse SFHs. The MIST models we use include stellar rotation that allows for the production of more ionizing photons at later times \citep{Choi2016}. Additionally, \prospector\ incorporates emission lines from \citet{Byler2017}.

Using FAST-like parameters (fixed $Z_* = Z_\odot/5$, delayed-$\tau$ SFHs) with our \prospector\ model (including MIST models), we recover properties for our EELGs matching the results in \citet{Forrest2017}. With our standard \prospector\ model, we find higher stellar masses for both EELGs and SFGs by $\sim 0.7$ dex and $\sim 0.5$ dex, respectively. We attribute this offset to the only two aspects of the model that we changed. First, the stellar metallicities (Z$_\odot / 50$) fit by \prospector\ are very low compared to the fixed Z$_\odot / 5$ used in FAST. Additionally, the SFR in older time bins in \prospector\ (see \S \ref{npsfh}) is poorly constrained. When the data are not constrained well, the priors dominate the posterior distributions, and the exponentially declining SFHs in FAST impose very different priors from the \lshort\ nonparametric SFHs.

\subsubsection{Nonparametric star formation histories\label{npsfh}}
Nonparametric SFHs are calculated in \prospector\ by fitting for fraction of total stellar mass formed in bins of lookback time, ranging from the time of the galaxy's observation to the age of the universe. Mock tests performed at lower redshifts indicate that \prospector\ accurately recovers summary statistics like the average age.  However, we note that  recovery of SFR in specific time bins is challenging (\lshort).

\begin{table}
	\begin{center}
	\textbf{Table 1} \\
	Nonparametric SFH bins \\
	\begin{tabular}{ |l|r| }
	\hline
	Bin name & \multicolumn{1}{|l|}{Lookback time} \\
	\hline
	\hline
    $dt1$ & $0-50$ Myr \\
    \hline
	$dt2$ & $50-100$ Myr \\
	\hline
    $dt3$ & $100$ Myr $- 1$ Gyr \\
	\hline
	$dt4$ - $dt6$ & Varies (see caption) \\
	\hline
	\end{tabular}
	\end{center}
	\caption{We tabulate the nonparametric SFH bins we use in \prospector, with the bin widths recorded in lookback time. $dt4$, $dt5$, and $dt6$ are evenly spaced from 1 Gyr to the age of the universe -- because our galaxies are at different redshifts, the exact width of these bins varies by galaxy. Note that $dt$6 is not a free parameter, as the fraction of star formation fit to the SFH bins must sum to 1 (see \lshort).}
	\label{sfh_table}
\end{table}
The edges of our time bins are chosen to isolate young, massive stars, which we expect to see in galaxies undergoing bursts of star formation. The youngest bin ($dt1$) runs from 0-50 Myr, followed by a 50 Myr-100 Myr bin ($dt2$), a 100 Myr-1 Gyr bin ($dt3$), and three bins evenly spaced from 1 Gyr to the age of the Universe at the redshift of the given galaxy ($dt4-6$). Because SFR is constant within each bin, the width of {$dt1$} puts a ceiling on the sSFR allowed in the model. We test and confirm that our results are robust to different choices of SFH bins.

The prior employed for our nonparametric SFH bins corresponds to constant star formation, i.e. a uniform prior. \lejaprep\ finds that with broadband photometry, our nonparametric SFHs clearly distinguish between various SFH shapes, including rising, falling, and bursty SFHs. \lejaprep\ also shows that this prior tends to over-estimate modestly the ages and stellar masses of galaxies with rising SFHs.

\subsection{Mock tests\label{mock}}
To determine the accuracy with which we can constrain our model parameters, we generate photometry for 100 mock EELG-like and 100 mock SFG-like galaxies. We randomly select from within the priors of our free parameters to create the mock galaxies, then use \prospector\ to generate the galaxy photometry in \zfourge\ filters. To simulate noise, we perturb the generated photometry by sampling from a Gaussian centered on zero, with a width equal to $5\%$ of the given flux at each point. The photometric errors are drawn randomly at each point to be between 0 and 10\% of the given flux, which is on the order of the errors in our real data. Additionally, we randomly select the field and redshift for each mock galaxy, choosing redshifts within $2.5 \leq z \leq 4$.

\begin{figure*}
\includegraphics[width=\textwidth]{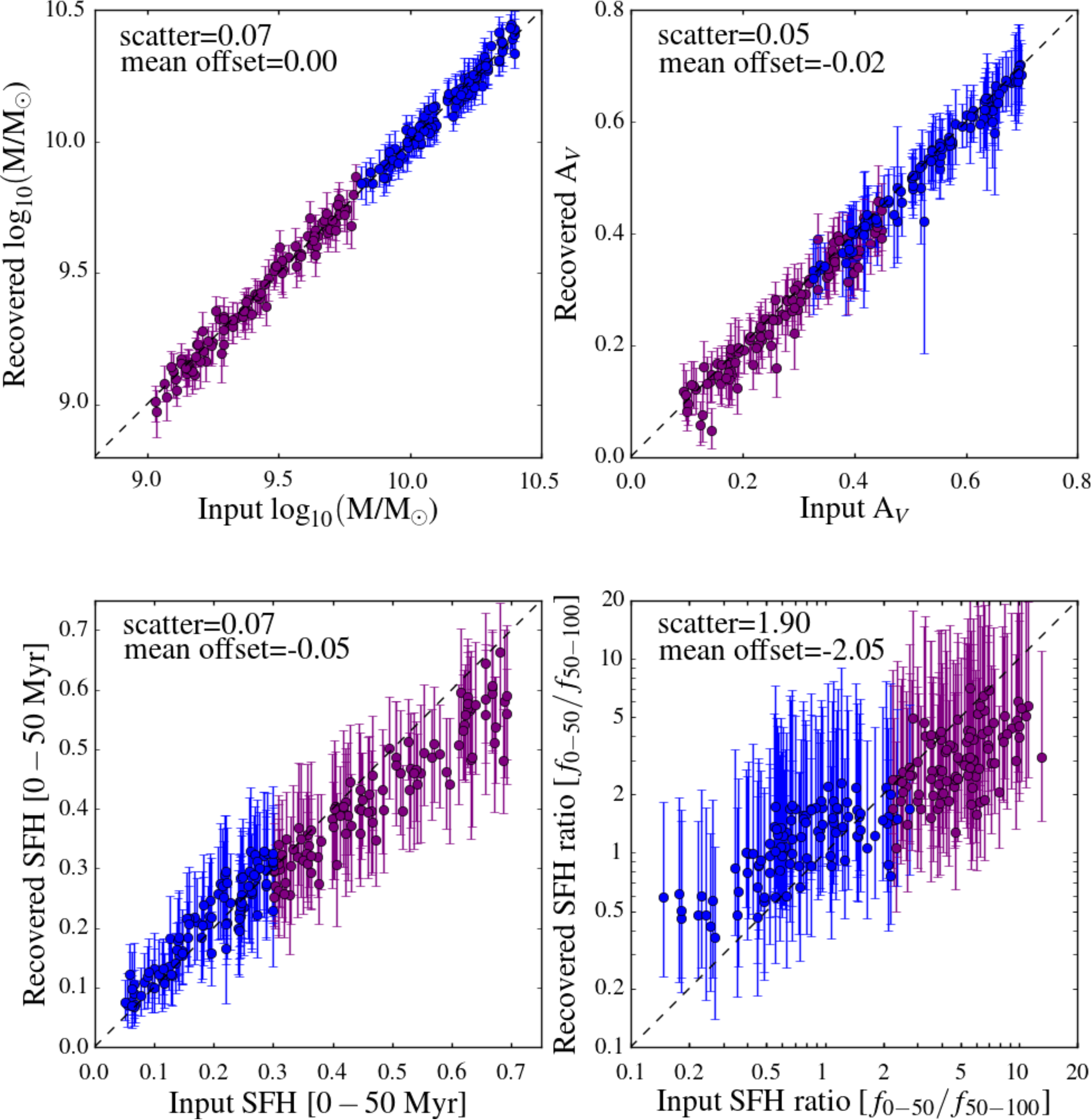}
\caption{Recovered vs input free parameters in \prospector\ for 100 EELG-like (purple) and 100 SFG-like (blue) mock galaxies with \zfourge-like photometry. We find that stellar mass, dust attenuation, and $dt1$ are well-recovered. Imperatively, we note that the ratio of $dt1$ ($0 - 50$ Myr) to $dt2$ ($50 - 100$ Myr) is also well-recovered. We conclude that, with \zfourge\ data, \prospector\ is capable of providing constraints on the SFHs of our EELGs and SFGs. Furthermore, correcting for the observed slope offset in the recovered vs input SFH ratio only strengthens the discrepancy between SFG and EELG-like SFHs (see Figure \ref{stack}).}
\label{recovery}
\end{figure*}

The results of the mock tests are displayed in Figure \ref{recovery}. We find that the stellar mass, dust, and $dt1$ are all well-recovered by \zfourge-like photometry. Crucially, the ratio between $dt1$ and $dt2$ is also well-recovered, indicating \prospector\ is capable of constraining SFHs from \zfourge-like photometry.

\section{\label{results}Results and Discussion}
\subsection{\label{starburst}EELGs show rising star formation histories}
\begin{table}
	\begin{center}
	\textbf{Table 2} \\
	Properties of composite SEDs \\
	\begin{tabular}{ |l|r|r| }
	\hline
	Property & \multicolumn{1}{|l|}{EELGs} & \multicolumn{1}{|l|}{SFGs} \\
	\hline
	\hline
    N$_{\text{galaxies}}$ & \numEELG & \numSFG \\
    \hline
    \hline
	sSFR [Gyr$^{-1}$] & $4.6^{+4.4 (+8.1)}_{-2.3 (-3.6)}$ & $1.1^{+1.6(+5.1)}_{-0.7(-1.1)}$ \\
	\hline
    log$_{10}$($\frac{\text{M}^*}{\text{M}_\odot}$) & $9.4^{+0.4(+0.5)}_{-0.4(-0.5)}$ & $10.1^{+0.3(+0.5)}_{-0.2(-0.5)}$ \\
	\hline
	$A_{\text{V}}$ & $0.32^{+0.14(+0.24)}_{-0.17(-0.26)}$ & $0.55^{+0.21(+0.36)}_{-0.22(-0.43)}$ \\
	\hline
	$\frac{\text{Z}_*}{\text{Z}_\odot}$ & $0.02^{+0.01(+0.34)}_{-0.00(-0.01)}$ & $0.02^{+0.04(+0.67)}_{-0.01(-0.01)}$ \\
	\hline
	K$_s$-band r$_{\text{h}}$ [kpc] & $0.84^{+0.10(+0.47)}_{-0.12(-0.17)}$ & $1.03^{+0.19(+0.53)}_{-0.17(-0.26)}$ \\ 
	\hline
	\end{tabular}
	\end{center}
	\caption{We tabulate the median, 16th percentile, and 84th percentile (95\% confidence intervals in parentheses) based on the population distributions (see \S \ref{starburst}) for sSFR, stellar mass, dust attenuation, and stellar metallicity derived from \prospector, and the K$_s$-band half-light radius determined with Source Extractor \citep{Bertin1996}. Our uncertainties are dominated by scatter in the distribution, rather than measurement error.}
	\label{table}
\end{table}

For each galaxy, we sample the posterior of each parameter $10^3$ times, building distributions for the EELGs and SFGs. We then calculate the median, 16th percentile, and 84th percentile for these distributions. From the output SFHs and stellar masses we compute specific SFHs (sSFHs) for each galaxy. We similarly calculate distributions of fractional mass formed in $dt1$ for each galaxy. The error on the mean for each of our parameter distributions is very small ($< 1\%$), indicating that our uncertainties are dominated by scatter in the distribution, rather than measurement error. The sSFR, stellar mass, dust attenuation, and stellar metallicity are recorded for each sample in Table \ref{table}.

\begin{figure*}
\includegraphics[width=\textwidth]{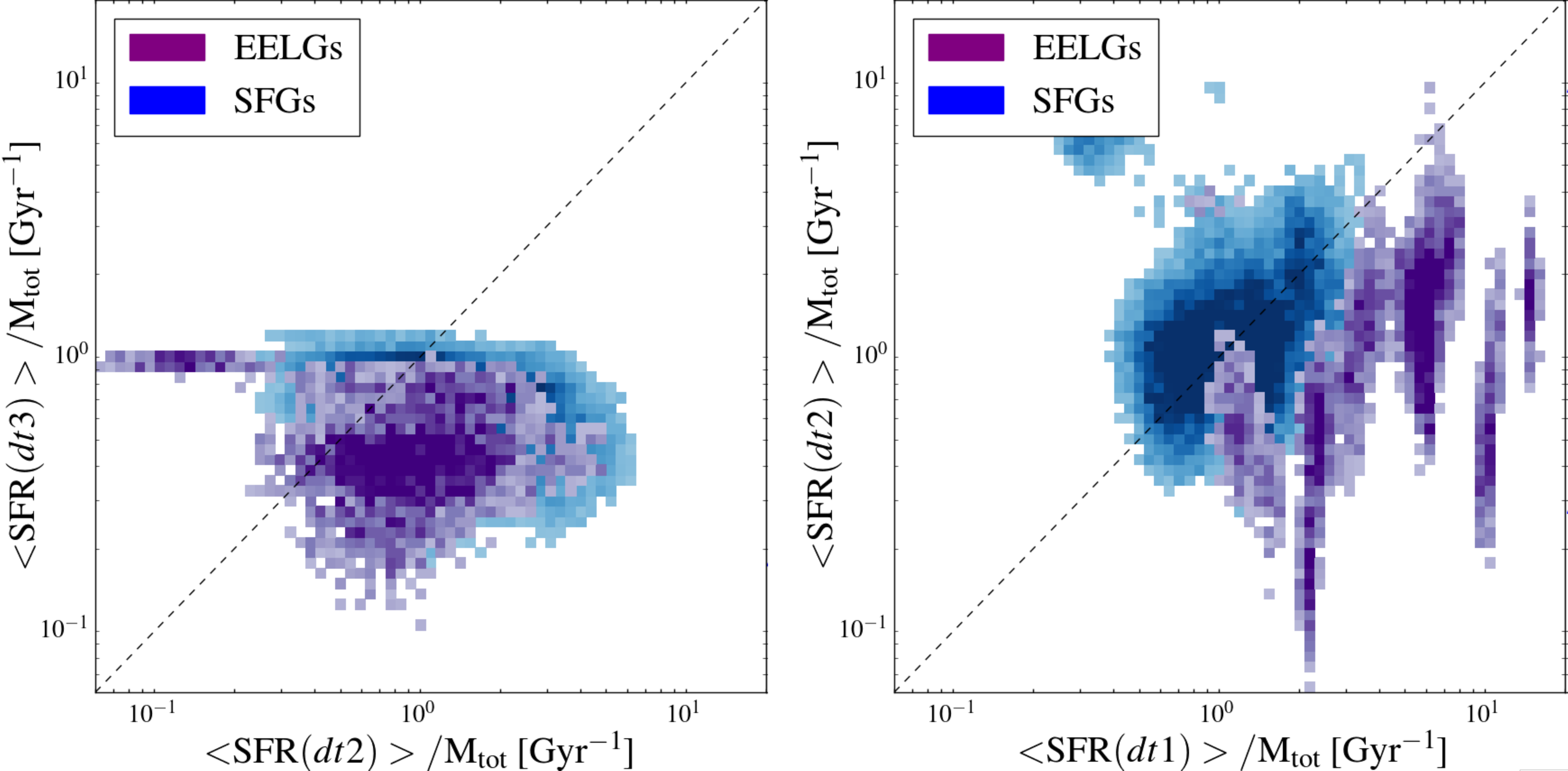}
\caption{Left: 2D histograms showing the average SFR in $dt3$ ($100-1000$ Myr; see Table \ref{sfh_table} for SFH bin definitions) divided by total stellar mass (M$_{\rm tot}$) vs the average SFR in $dt2$ ($50-100$ Myr) divided by M$_{\rm tot}$. Right: 2D histogram showing the average SFR in $dt2$ ($50-100$ Myr) divided by M$_{\rm tot}$ vs the average SFR in $dt1$ ($0-50$ Myr) divided by M$_{\rm tot}$. We draw randomly $10^3$ times from the SFR posterior distributions of each galaxy in each sample. In each panel, the resulting EELG distribution is shown in purple and the SFG distribution in blue. Both populations show greater SFRs in $dt2$ than in $dt3$ (left panel). However, the EELG population shows an increase in star formation from $dt2$ to $dt1$ (right panel), and we conclude that the EELGs display rising SFHs across both bins. In contrast, the SFGs are centered on the 1:1 (dashed) line in the right panel, showing no preference for rising SFHs.}
\label{stack}
\end{figure*}
The distributions of SFR normalized by total stellar mass, both in $dt1$ ($0-50$ Myr) and $dt2$ ($50-100$ Myr), is displayed for each sample in the right panel of Figure \ref{stack}. EELGs are twice as likely to have a recent burst compared to SFGs ($\sim 90\%$ vs $\sim 40\%$). Similarly, we display the distributions in $dt2$ ($50-100$ Myr) vs $dt3$ ($100-1000$ Myr) in the left panel of Figure \ref{stack}. The EELGs and SFGs both show rising SFRs on average ($\sim 70\%$ vs $\sim 80\%$) from $dt3$ to $dt2$, but only the EELGs show rising SFRs from $dt2$ to $dt1$, indicating the EELGs have rising SFHs.

With these SFHs, the EELGs display significant star formation only in the past 50 Myr, suggesting that this most recent episode is a dominant starburst. By comparison, most of the SFGs have more constant SFHs. The EELGs and SFGs show median sSFRs of 4.6 Gyr$^{-1}$ and 1.1 Gyr$^{-1}$, respectively. We hypothesize that the steep UV slopes and large EWs associated with our EELGs (see e.g. Figures \ref{seds} and \ref{ratio}) are driven by their rising SFRs.

Whether the EELGs maintain or decrease their SFRs, their specific SFRs will decrease due to their increasing stellar masses. As such, we hypothesize that the current EELG sSFR distribution will evolve to look more like that of the SFGs. Given the mass, dust, and sSFR offsets between the EELG and SFG samples, we hypothesize that our EELGs are plausible analogs for higher redshift progenitors of the more typical SFGs (e.g. \citet{Smit2015}).

\subsection{\label{other}Bursty vs constant star formation histories}
Within 68\% confidence intervals, the EELGs form $15^{+30}_{-10}\%$ of their stellar mass in the most recent 50 Myr, while the SFGs form only $4^{+9}_{-3}\%$ of their mass in the same time. As above, these uncertainties correspond to scatter in each distribution rather than measurement error. As with sSFR and dust (see \S \ref{starburst}), the offset in fractional mass formed remains when considering a mass-matched subsample of our EELGs and SFGs, though a larger sample is necessary for a more rigorous analysis.

We note that three of our EELGs form $\geq 60\%$ of their stellar mass in the most recent 50 Myr, with the strongest EELG forming $87\%$ of its mass. We hypothesize that such EELGs may be undergoing an initial major episode of star formation and designate them as candidate \textit{first burst galaxies}. However, the existing data do not allow us to test for the presence of a faint, older, red stellar population. Only with the resolution and wavelength range of e.g. JWST can this be tested.

The substantial fractional stellar mass formed by our EELG sample indicates many galaxies experience burst-dominated buildups of stellar mass. Our SFGs, as well as the EELGs with less fractional stellar mass formed in the last 50 Myr, may have undergone one or more such episode(s) of star formation prior to their current burst. However, we cannot rule out SFHs with a constant buildup of stellar mass for these galaxies.

We note that recent simulations, such as Feedback In Realistic Environments (FIRE; e.g. \citealt{Faucher-Giguere2018}) and FirstLight \citep{Ceverino2018}, suggest that star formation at high redshift is burst-dominated. Our observations are consistent with this picture, indicating that the disparities we see in EELG SEDs compared to SFG SEDs likely correspond to variations in recent SFH.

\subsection{\label{fraction}Fraction of galaxies with extreme emission}
To estimate the fraction of galaxies that go through an extreme \oh\ emission line phase between $2.5 \leq z \leq 4$, we compare the number of galaxies identified in emission-line galaxy composite SEDs in \fshort\ to the expected number of total galaxies based on the galaxy stellar mass function. We use the best-fit double-Schechter parameters from \citet{Tomczak2014} for star forming galaxies at $2.5 \leq z \leq 3$ to calculate the number density of star forming galaxies in the same mass range as our EELGs.

We find that $\gtrsim 2\%$ of galaxies with masses $\geq 10^9$M$_\odot$ at $2.5 \leq z \leq 4$ are undergoing extreme emission at the time of observation. This fraction is a lower limit, as the \zfourge\ survey is not complete at the masses of our EELGs. Assuming extreme emission timescales of $100 - 30$ Myr \citep[e.g.][]{Wel2011,Ceverino2018} implies that $\geq 20 - 67\%$ of star-forming galaxies go through an EELG phase at $2.5 \leq z \leq 4$. As such, we hypothesize that a large fraction, and possibly a majority, of low-mass galaxies go through an EELG phase at $z \geq 2.5$.

\subsection{\label{others}Metallicities, masses, and caveats}
As shown in Table \ref{table}, the median EELG has lower stellar mass and is offset to a higher sSFR, less dust, and a more compact size than the median SFG. We stress that we recover similar values for all parameters, including the strikingly low stellar metallicities, when we mask the emission lines in our \prospector\ fits.

We note that a subset of our EELGs and SFGs overlap in stellar mass at $9.7 \leq \log_{10}$(M / M$_\odot) \leq 9.9$. The \prospector-derived properties for the mass-matched EELG and SFG subsamples are reported in Table \ref{match}. Like the full samples, the mass-matched EELGs are offset to higher sSFRs and less dust compared to the mass-matched SFGs. These results indicate that the sSFR and dust differences we see between the EELGs and SFGs are not simply driven by the stellar mass offset.

\begin{table}
	\begin{center}
	\textbf{Table 3} \\
	Properties of mass-matched subsamples \\
	\begin{tabular}{ |l|r|r| }
	\hline
	Property & \multicolumn{1}{|l|}{EELGs (matched)} & \multicolumn{1}{|l|}{SFGs (matched)} \\
	\hline
	\hline
    N$_{\text{galaxies}}$ & 5 & 22 \\
    \hline
    \hline
	sSFR [Gyr$^{-1}$] & $2.8^{+0.3}_{-0.8}$ & $0.7^{+1.8}_{-0.4}$ \\
	\hline
    log$_{10}$($\frac{\text{M}^*}{\text{M}_\odot}$) & $9.8^{+0.1}_{-0.0}$ & $9.8^{+0.1}_{-0.0}$ \\
	\hline
	$A_{\text{V}}$ & $0.28^{+0.03}_{-0.09}$ & $0.40^{+0.24}_{-0.24}$ \\
	\hline
	$\frac{\text{Z}_*}{\text{Z}_\odot}$ & $0.03^{+0.22}_{-0.00}$ & $0.02^{+0.03}_{-0.01}$ \\
	\hline
	\end{tabular}
	\end{center}
	\caption{Here we tabulate the median, 16th percentile, and 84th percentile (based on the population distributions; see \S \ref{starburst}) for sSFR, stellar mass, dust attenuation, and stellar metallicity derived from \prospector\ for the mass-matched EELG and SFG subsamples. Like the full samples, the EELG mass-matched subsample is offset toward higher sSFRs and less dust attenuation compared to the SFG mass-matched subsample.}
	\label{match}
\end{table}

The stellar metallicity that we derive is lower than expected based on the metallicities of similar-mass galaxies in the $z=0$ Universe. This may indicate that the stellar metallicity value \prospector\ measures is a combination of the true metallicity with possible systematics or uncertainties in the modeling.

To investigate the effects of the low stellar metallicities on our results, we reduce the range of the stellar metallicity in our model. Using a lower bound of $Z_\odot / 10$, the fit stellar masses decrease by $\sim 0.2$ dex (as expected from \S \ref{fast}) and correspondingly return sSFRs a factor of 1.4 times larger than those in Table \ref{table}. However, our qualitative results regarding the recent burst of star formation in EELGs, the rising SFHs in our EELGs, and the differences in SFH between our EELG and SFG samples remain unchanged.

Comparing the reduced $\chi^2$ for tests with higher stellar metallicity floors, we rule out stellar metallicities on the order of $Z_\odot$ for our EELGs and SFGs. Although we cannot rule out stellar metallicities $\sim Z_\odot /10$ for individual galaxies, we find that the average population for both EELGs and SFGs shows a weak preference ($\Delta\chi^2 = 0.07^{+0.02}_{-0.02}$) for the lower stellar metallicities fit with a floor of $Z_\odot / 100$.

It is plausible that the extreme nebular conditions in our galaxies may be straining the model, affecting parameters like stellar metallicity. One other known inaccuracy is the geometry of the dust, the young stars, and the old stars, which are more complex in real galaxies than in our models. Our \citet{Calzetti1994} attenuation law may also lead to erroneous metallicity estimates. Systematically investigating these possibilities is beyond the scope of the present paper; here we simply caution that the fit parameters may not only reflect the physical parameters that we associate with them.

\section{\label{conclusions}Conclusions}
We explore the properties of \numEELG\ extreme \oh\ emission line galaxies and a population of more typical star-forming galaxies identified in the \zfourge\ survey. The EELGs are selected from the bluest composite SED built in \citet{Forrest2018} at $2.5 \leq z \leq 4$, and the SFGs are selected from the composite SED with the most analog galaxies at the same redshift range. In a companion paper, \ptran\ spectroscopically confirm 4 of our EELGs with the {\sc MOSEL} survey and find that the spectroscopically-confirmed EELGs trace the same property distributions as the photometrically-selected sample in \fshort.

We use the \prospector\ SED fitting code, which includes a physical model for nebular line and continuum emission, to estimate probability distribution functions for stellar mass, stellar metallicity, dust attenuation, and SFH for each of our galaxies. We compare the EELG and SFG distributions for each of these parameters. Compared to the SFGs, the EELGs have lower stellar masses and are offset toward higher sSFRs, less dust, and smaller half-light radii.

The EELGs display a dominant starburst in the most recent 50 Myr and are twice as likely to have a recent burst of star formation compared to SFGs ($\sim 90\%$ vs $\sim 40\%$). Furthermore, the EELGs form $15^{+30}_{-10}\%$ of their stellar mass in the most recent 50 Myr, with a subset forming $\geq 60\%$ of their stellar mass; we coin the latter \textit{first burst galaxies}. In contrast, the SFGs form only $4^{+9}_{-3}\%$ of their stellar mass in the same time. The median sSFR for the EELGs is 4.6 Gyr$^{-1}$ and for the SFGs is 1.1 Gyr$^{-1}$.

The EELG and SFG stellar masses from \prospector\ are offset to higher values than earlier results e.g. with FAST. This is likely due to the lower stellar metallicities and nonparametric SFHs. Additionally, our EELGs and SFGs have sub-solar stellar metallicites, but we note that our model does not strongly constrain the true value. However, we find that applying different stellar metallicity floors does not change our qualitative SFH results. Future work involving improved modeling of the UV slope and reducing uncertainties in the youngest stellar populations (e.g. $<10$ Myr) are needed to better constrain the properties of EELGs and SFGs at $z>2$.

From the number density of EELGs, we estimate that at least $20\%$ of star-forming galaxies with stellar masses $\geq 10^9$M$_\odot$ display extreme emission between $2.5 \leq z \leq 4$. Our results indicate that burst-dominated growth may be common at $z \geq 2.5$ and that these episodes can drive extreme emission lines. The key next step in characterizing the relationship of the EELGs to the larger galaxy population is establishing the typical duration of the EELG phase.

New, groundbreaking telescopes such as JWST will be able to detect H$\alpha$ emission from these galaxies and provide more reliable measurements of SFR, stellar metallicity, the ionization parameter, and electron temperature. More accurate estimates for these parameters will help test scaling relations at high redshifts, improving our SED modeling techniques by formulating more informative priors. Additionally, calculating gas masses (e.g. with ALMA) would be instrumental in estimating SFH burst timescales. Follow-up observations with such telescopes will thus provide more insight into the early lives of SFGs, EELGs, and the prevalence of \oh\ extreme emission as an early-lifetime phase.

\section*{Acknowledgements}
We wish to thank the Mitchell family for their continuing support of astronomy. We also thank the Carnegie Observatories and the Las Campanas Observatory for their assistance in making the \zfourge\ survey possible. K. Tran thanks P. Oesch, B. Holden, and M. Maseda for helpful discussions and acknowledges support by the National Science Foundation under grant \#1410728. GGK acknowledges the support of the Australian Research Council through the award of a Future Fellowship (FT140100933). BF would like to thank the Hagler Institute for Advanced Study at Texas A\&M. JHC thanks T. Hutchison for constructive comments on the first draft. Portions of this research were conducted with high performance research computing resources provided by Texas A\&M University (\url{https://hprc.tamu.edu}). We also thank the anonymous referee for valuable comments that improved the manuscript.

\end{document}